\documentclass[namedreferences]{solarphysics}
\usepackage[hyperref,optionalrh,solaromanenum]{spr-sola-addons} % For Solar Physics 
\usepackage[dvipdfmx]{graphicx}                    % For eps figures, newer & more powerfull
\usepackage[dvipdfmx]{color}
\usepackage{color}                       % For color text: \color command
\newcommand{\aap}{    {\it Astron. Astrophys.}}
\newcommand{\apj}{    {\it Astrophys. J.}}
\newcommand{\apss}{   {\it Astrophys. Space Sci.}} 
\newcommand{\pasa}{   {\it Pub. Astron. Soc. Australia}}
\newcommand{\pasj}{   {\it Pub. Astron. Soc. Japan}}
\newcommand{\pasp}{   {\it Pub. Astron. Soc. Pac.}}
\newcommand{\solphys}{{\it Solar Phys.}}
\newcommand{\ssr}{    {\it Space Sci. Rev.}} 

\begin{document}

\begin{article}

\begin{opening}

\title{Automated Sunspot Detection as an Alternative to Visual Observations}

%%%%%%%%%%%%%%%%%%%%%%%%%%%%%%%%%%%%%%%%%%%%%%%%%%%
%% Authors Names
%
\author[addressref={aff1},corref,email={yoichiro.hanaoka@nao.ac.jp}]{\inits{Y.H.}\fnm{Yoichiro}~\lnm{Hanaoka}\orcid{0000-0003-3964-1481}}

%%%%%%%%%%%%%%%%%%%%%%%%%%%%%%%%%%%%%%%%%%%%%%%%%%%
%% Runningheads
%
\runningauthor{Y. Hanaoka}
\runningtitle{Automated Sunspot Detection}

%%%%%%%%%%%%%%%%%%%%%%%%%%%%%%%%%%%%%%%%%%%%%%%%%%%
%% Affilations 
%% id shold be the same with \author addressref value.
\address[id={aff1}]{National Astronomical Observatory of Japan, Mitaka, Tokyo 181-8588, Japan}

%%%%%%%%%%%%%%%%%%%%%%%%%%%%%%%%%%%%%%%%%%%%%%%%%%%
%%% Abstract 
\begin{abstract}
We developed an automated method for sunspot detection using digital white-light solar images to achieve a performance similar to that of visual drawing observations in sunspot counting.
To identify down to small, isolated spots correctly, we pay special attention to the accurate derivation of the quiet-disk component of the Sun, which is used as a reference to identify sunspots using a threshold.
This threshold is determined using an adaptive method to process images obtained under various conditions. 
To eliminate the seeing effect, our method can process multiple images taken within a short time.
We applied the developed method to digital images captured at three sites and compared the detection results with those of visual observations. We conclude that the proposed sunspot detection method has a similar performance to that of visual observation. 
This method can be widely used by public observatories and amateurs as well as professional observatories as an alternative to hand-drawn visual observation for sunspot counting.

\end{abstract}

%%%%%%%%%%%%%%%%%%%%%%%%%%%%%%%%%%%%%%%%%%%%%%%%%%%
%% Keywords
%

\end{opening}
%-------------------------------------------------

%%%%%%%%%%%%%%%%%%%%%%%%%%%%%%%%%%%%%%%%%%%%%%%%%%%
%% Sections
%
\section{Introduction}\label{s:1} 

Sunspots have been observed using telescopes for more than 400 years. Well-calibrated sunspot numbers comprehensively encode the variation in solar activity during that period \citep{2014SSRv..186...35C}. 
Generally, the relative sunspot number is expressed as $k(10g+f)$, where $f$, $g$, and $k$ are the number of sunspots, number of sunspot groups, and the correction factor, which depends on the observer and/or instrument, respectively. The relative sunspot number is used as a simple index of the solar activity based on sunspot counts.
Persistence and simplicity are the reasons why the sunspot number is still an important index of the solar activity, even though various more quantitative indices, such as the area of sunspots and their magnetic field flux, are now available. 

The relative sunspot number, currently known as the international sunspot number, which is the relative sunspot number now maintained by the Sunspot Index and Long-term Solar Observations (SILSO) of the Royal Observatory of Belgium, is still based on sunspot counts on hand-drawn sketches obtained by visual observations. Not only the pilot station, i.e., Specola Solare Ticinese at Locarno, Switzerland \citep{2016SoPh..291.3075C}, and some principal observatories, but many observers worldwide provide sunspot-count data mainly obtained by visual observations of the Sun. However, it is becoming increasingly difficult to secure manpower to carry out such an old-style observation, particularly for professional and public observatories. 

Meanwhile, acquisition of white-light full-disk digital images of the Sun is being regularly carried out by many observatories and some spacecraft. Automated detection of sunspots on digital white-light images enables objective sunspot counting and allows regular sunspot observations with small manpower.

For this reason, the National Astronomical Observatory of Japan (NAOJ) changed the sunspot counting method from hand-drawn visual observation to automated sunspot detection using digital white-light images taken with a 10-cm refractor and a CCD camera (with a resolution of approximately 2000$\times$2000 pixels) in 1998 \citep{1998RNAOJ...4....1I, 2002AdSpR..29.1565S}. However, detections of false spots and missed detections of true spots often occur. The performance of this sunspot detection is not comparable to that of visual observations. Therefore, we developed another high-performance method and applied it to higher-quality white-light data to improve the performance of sunspot detection. If software for such a method can process data taken with various instruments, it is expected to contribute to the realization of modernized sunspot counting observations by observers other than the NAOJ.

In recent years, various automated techniques for detecting features in solar images have been developed \citep[see e.g.,][]{2010SoPh..262..235A}, and many studies have been conducted on sunspot detection \citep[see e.g.,][]{2018PASP..130j4503Y}. Various techniques have been proposed for sunspot detection using full-disk images. The thresholding technique and corresponding modifications, through which structures darker than a certain threshold are identified as sunspots, were used by \cite{2001SoPh..202...53P}, \cite{2008SoPh..250..411C}, \cite{2008SoPh..248..277C}, \cite{2009SoPh..260....5W}, \cite{2016PASA...33...18Z}, and \cite{2018PASP..130j4503Y} for data acquired by San Fernando Observatory, Ebro Observatory, Solar and Heliospheric Observatory/Michelson Doppler Imager \citep[SOHO/MDI;][]{1995SoPh..162..129S}, again MDI, Huairou Solar Observing Station, and Solar Dynamics Observatory/Helioseismic Magnetic Imager \cite[SDO/HMI;][]{2012SoPh..275..207S}, respectively. In \cite{2020A&C....3200385C}, results using thresholding and mathematical morphological operations were compared on data from the Coimbra Observatory. MDI data were also processed in \cite{2002ApJ...568..396T} using the Bayesian image-segmentation technique, in \cite{2005SoPh..228..377Z} using edge detection, and in \cite{2014SoPh..289.1413G} using level-set image-segmentation.

Although a great deal of methods for sunspot detection have already been developed, many of them are not necessarily adequate for sunspot counting. Their main purpose is to obtain the area of sunspots because it is more objective than the relative sunspot number when it comes to representing sunspot activity. Existing methods are not necessarily prepared to detect small spots, which make little difference to the total area. By contrast, drawing observation has an important unique feature in that special attention is paid to isolated small spots. An isolated pore forming a sunspot group alone yields a difference of 11 ($10g+f$ with $g=f=1$) in the relative sunspot number. Although it is not quantitatively correct that the solar activity index significantly depends on the existence of an isolated pore, to continue performing sunspot counting in the same traditional manner but with automated methods, it is important to capture isolated pores to the same degree as visual observations. Although the difference in the detection performance of small spots can be compensated for by the factor $k$, observations with too many missed detections of small spots are difficult to consider statistically meaningful. 
Nevertheless, missed detections of spots that are barely captured by high-quality visual observations is allowed to some extent. Even in this case, the missed spots should not exhibit uneven distribution on the disk. An uneven distribution means, for example, a bias in the frequency of missed detections depending on the distance from the disk center.

Furthermore, most previously proposed methods aim to process a specific dataset. In particular, many studies have been conducted to process data taken with spacecraft without seeing effect, but their flexibility is low. It is desirable to develop a method that can be used by public observatories and individual amateurs who carry out white-light imaging observations of the Sun. Data from such observations exhibit variations resulting from the telescope, number of pixels, and bit depth of the solar image. Additionally, the seeing conditions change over time. Flexibility covering such varieties is required for automated sunspot detection.

Moreover, another capacity is desirable for automated sunspot detection. In former automated methods, a single image was usually processed. However, there is an essential difficulty in sunspot detection using a single image taken from ground-based observation. The seeing effect sometimes produces short-lived non-sunspot dark structures that are difficult to distinguish from true sunspots. In visual observations, observers watch the variation in the visibility of small dark features and judge whether they are true spots. This has been recognized as an advantage of visual observations.
However, if a series of digital images are taken within a short time, a false dark spot appearing in one of the images disappears in almost all other images. In contrast, true sunspots appear in most of the images in a series. Processing multiple images is expected to increase the reliability of sunspot detection using digital images. Multiple-image acquisition is now carried out by some observatories and amateurs, and such data are commonly available.

Consequently, to achieve a sunspot counting performance comparable to that of visual drawing observation, we developed a new automated sunspot detection method. 
In Section~\ref{s:2}, we describe the method developed for detecting sunspots. The results of the application of this method to white-light observations and its performance evaluation are presented in Section~\ref{s:3}. The conclusions are summarized in Section~\ref{s:4}.

\section{Detection Method}\label{s:2} 

\subsection{Key Requirements for the Detection Method}\label{s:2-1} 

The key requirements for the proposed automated sunspot detection method, which has a performance comparable to that of visual drawing observation, are summarized next.

\begin{itemize}
\item It should detect sunspots as visual drawing observations do. In other words,
\begin{itemize}
\item The number of sunspots detected with the automated method should be comparable to that of visual observations if the quality of digital images is sufficiently high.
\item False detection of sunspots should be as low as in visual observations.
\item Missed detections of true spots are allowed to some extent, but they should not exhibit a biased distribution.
\end{itemize}
\item It should be flexible enough to accept digital data taken at a variety of conditions.
\item It should process multiple images taken within a short time altogether to eliminate erroneous detections due to the seeing effect.
\end{itemize}
The developed detection method was designed to fulfill these requirements. 

There are two basic procedures for automated sunspot detection by using a single image. The first step is intensity normalization to produce a contrast image such that the original full-disk image is converted to a relative brightness distribution with respect to the quiet region outside the sunspots. 

The second step involves identifying sunspots in the contrast image. Among the various techniques used to identify sunspots, we adopted the threshold technique, in which areas darker than a certain threshold in contrast images are identified as sunspots. Although the primary purpose of sunspot detection is to count the number of sunspots, the data resulting from sunspot detection are expected to be used for quantitative analysis, such as evaluation of sunspot areas and deficit of light flux due to sunspots. In the proposed threshold method, the quantitative characteristics of the detected spots are unambiguously determined using an explicit criterion.

For multiple-image data, an additional procedure to determine valid spots using multiple images is applied after applying the two above steps to each image.

In the following subsections, we describe these procedures using data taken with the Solar Flare Telescope (SFT) of the NAOJ (see Section~\ref{s:3-1} for further details) on 2014 February 28 as an example.

\subsection{Intensity Normalization}\label{s:2-2} 

% Figure 1
\begin{figure} 
\centerline{\includegraphics[width=1\textwidth]{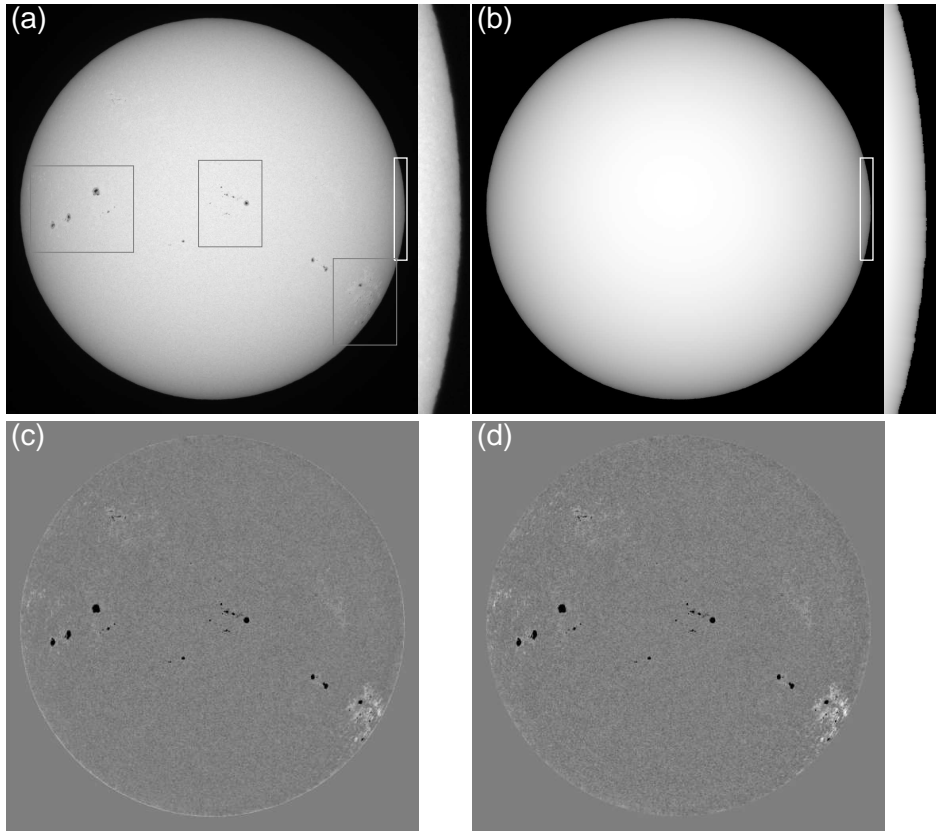}}
\caption{Images illustrating various steps of the proposed automated sunspot detection. (a) An original white-light image of the Sun on 2014 February 28 taken with the SFT. An enlargement of the box marked with a white frame is shown to the right to present the blurring caused by the seeing effect. Detected sunspots in three gray frames are depicted in Figure~\ref{fig:6}. (b) Disk component without sunspots. An enlargement of a part of the limb is shown to the right. (c) Contrast image after normalization using the disk component. (d) Final contrast image for sunspot identification. Limbward enhancement of the contrast and removal of erroneous pixels at the limb were applied. In all panels, the celestial north is to the top, and the solar north is rotated clockwise by 21.25 degrees from the celestial north.}\label{fig:1}
\end{figure}

Next, we explain the first step of the sunspot detection process, that is, intensity normalization. An example is shown in Figure~\ref{fig:1}.
Although a specific image is used in the explanation below, images with various numbers of pixels, bit depths, and qualities can be processed in similar terms.

Starting from an original 2080$\times$2080-pixel white-light image (Figure~\ref{fig:1}(a)), we estimated the brightness distribution of the quiet disk component without sunspots (Figure~\ref{fig:1}(b)), and normalized the original image with it (Figure~\ref{fig:1}(c)). Figure~\ref{fig:1}(d) depicts the result of additional corrections. The final image shows the relative brightness of the structures on the solar disk, such as sunspots. Note that sunspots can be identified simply by inspecting their darkness in the final contrast image. 

To produce a quiet-disk component without sunspots, we fit the observed limb darkening with a single curve, as done in many studies \citep[e.g.,][]{2003SoPh..214...89Z}. 
Apart from such fitting, mathematical morphological operations, particularly closing operations, have been used to remove sunspots and derive spotless disk components \citep{2008SoPh..250..411C, 2009SoPh..260....5W, 2016PASA...33...18Z}. However, these operations are not appropriate for detecting sunspots with high accuracy. The quiet solar disk exhibits brightness fluctuations due to granulation, and the closing operation fills dips between granulation cells as well as those due to sunspots. The brightness of the disk produced by the closing operation represents the brightest part of granulation (Figure 1 in \cite{2009SoPh..260....5W} illustrates this phenomenon). However, the brightness (darkness) of sunspots should be expressed by referring to the average granulation brightness; therefore, the closing operation introduces errors in the sunspot brightness. Additionally, \cite{2018PASP..130j4503Y} pointed out that the removal of the disk component using a closing operation causes unnecessary smoothing of sunspot images, and sometimes small dark features are deleted. 

% Figure 2
\begin{figure} 
\centerline{\includegraphics[width=0.7\textwidth]{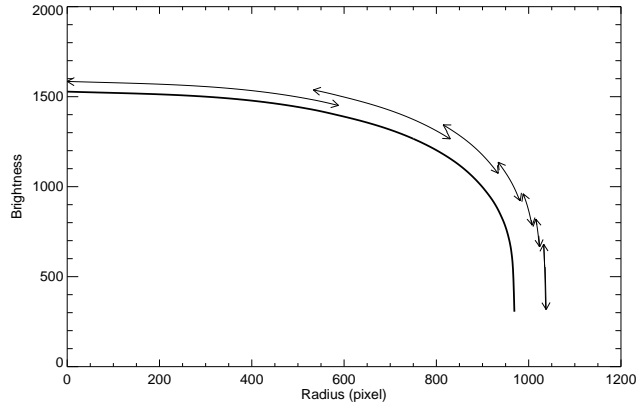}}
\caption{Fitted limb-darkening curve for the image shown in Figure~\ref{fig:1}(a). The ranges depicted by two-headed arrows were separately fitted with individual cubic polynomials.}\label{fig:2}
\end{figure}

Therefore, we adopted fitting of the limb-darkening curve to reproduce the quiet-disk component. 
This basic process is not different from that discussed in \cite{2003SoPh..214...89Z}. First, the limb positionand the center and radius of the disk are derived. Then, an average limb-darkening curve is derived based on the median brightness at each radial position.

A single polynomial function fits well with true limb darkening \cite[e.g.,][]{1977SoPh...51...25P}, but the observed profile is often distorted by the seeing effect and nonlinear response to brightness signals. Therefore, we adopted a more flexible fitting method. Figure~\ref{fig:2} shows the fitting results for the image in Figure~\ref{fig:1}(a). The solar radius was divided into seven sections, indicated by the two-headed arrows in Figure~\ref{fig:2}, and the observed profile in each section was fitted with a cubic polynomial. An overlap between two adjacent sections resulted from this process; interpolation of cubic polynomials was applied to the overlapping parts.

However, a simple, circularly symmetrical disk produced from the fitted limb-darkening curve is not necessarily adequate for high-precision sunspot detection, which requires a correctly derived disk brightness around each sunspot. The disk component suffers from distortion by the seeing effect and exhibits a nonuniform brightness distribution; therefore, the quiet-disk component should reproduce these effects as well. 
The modification of the symmetrical disk component is explained next. 

A large spatial-scale discrepancy between the reproduced disk from the limb-darkening curve and the observed results usually occurs due to incomplete (or no) flat fielding. These components are evaluated and added to the reproduced disk.

Furthermore, the observed disk is affected by the seeing effect, and the observed limb undulates, as shown in the enlarged image in Figure~\ref{fig:1}(a). Therefore, the difference between the reproduced disk and observed results sometimes shows remarkable discrepancies near the limb. To reproduce the observed limb shape, we stretched the synthesized disk to adapt to the undulating limb. The enlarged image in Figure~\ref{fig:1}(b) depicts a part of the limb adapted to the observed limb shown in Figure~\ref{fig:1}(a).

In \cite{2020A&C....3200385C}, the authors pointed out that the brightness distribution of the disk derived by the morphological operation is better for detection of spots near the limb than that based on fitting of the limb darkening. The reason is that the morphological operation performs well with a distorted disk. However, our method for reproducing the disk with modifications using the shape of the limb eliminates this disadvantage in limb-darkening fitting.

By dividing the observed image by the reproduced disk, we obtain a contrast image in which the sunspots are seen as depressions from a constant background level, as shown in Figure~\ref{fig:1}(c). In automated detection, depressions with a brightness below a certain threshold are identified as sunspots. However, we found that such identified sunspots are not consistent with those found by visual inspection of the image; spots near the limb tend to be missed in automated detection. The contrast of granulation is maximized around the disk center and decreases towards the limb. Therefore, small spots are more easily identified visually towards the limb. To accommodate this tendency, the contrast is enhanced by compensating for the radial decrease in the average contrast of the granulation. 

This enhancement sometimes unnecessarily increases the contrast of pixels very close to the limb. At a certain elongation from the disk center, the frequency of such error pixels (which are defined as having a contrast exceeding 5 \%) reaches its maximum. The pixels at this elongation or farther are excluded from sunspot detection.
Figure~\ref{fig:1}(d) shows the results of contrast enhancement and rejection of error pixels based on Figure~\ref{fig:1}(c).

\subsection{Sunspot Detection on a Contrast Image}\label{s:2-3} 

In the final contrast image shown in Figure~\ref{fig:1}(d), the ordinary photosphere and penumbrae are discriminated by a certain threshold. Some of penumbrae include umbrae, and umbrae are discriminated using another threshold. Determining a proper threshold for penumbrae is critical in automated sunspot detection because a proper threshold prevents missing true sunspots and detection of false spots. 
Therefore, next we mainly discuss the determination of the threshold for penumbrae.

Regarding the threshold for umbrae, $T_u$, the depression of about 30—40 \% was derived by previous studies, and it is 3—4 times of the threshold for penumbrae, $T_p$ \citep{1990Ap&SS.170..127S, 1996A&A...310..635S, 1993SoPh..146...49B, 1994svsp.coll..117C}. Based on these results and considering the fact that $T_u$ depends on the seeing similarly to $T_p$, we defined that $T_u = 3 \times T_p -0.05$ (the lower limit is $-0.41$, which means a depression of 41 \% from neighboring photosphere). We confirmed that such derived $T_u$ values generally give consistent discrimination results with those done visually. However, in reality, the brightness inside of penumbrae is affected by scattered light, and therefore, the brightness at the boundary between penumbrae and umbrae often depends on the size of penumbrae. Although a common threshold for all of the umbrae is necessary for quantitative image analysis, it should be noted that it is not necessarily suitable for individual umbrae.

To determine the proper threshold for penumbrae, fixed thresholds or standard deviation of the brightness in images multiplied by fixed factors have often been used in previous studies, as reviewed by \cite{2018PASP..130j4503Y}. However, we aim to process images taken not only under various seeing conditions but also with a variety of observing systems. Therefore, we adopted a more flexible method for determining the threshold. We try to detect penumbrae with various thresholds and select the threshold value that provides the most appropriate discrimination of penumbrae as the best threshold, as done by \cite{2005SoPh..228..377Z} and \cite{2008SoPh..250..411C}. In \cite{2018PASP..130j4503Y}, another self-adaptive threshold-determination method using artificial intelligence technology was proposed.

% Figure 3
\begin{figure} 
\centerline{\includegraphics[width=1.\textwidth]{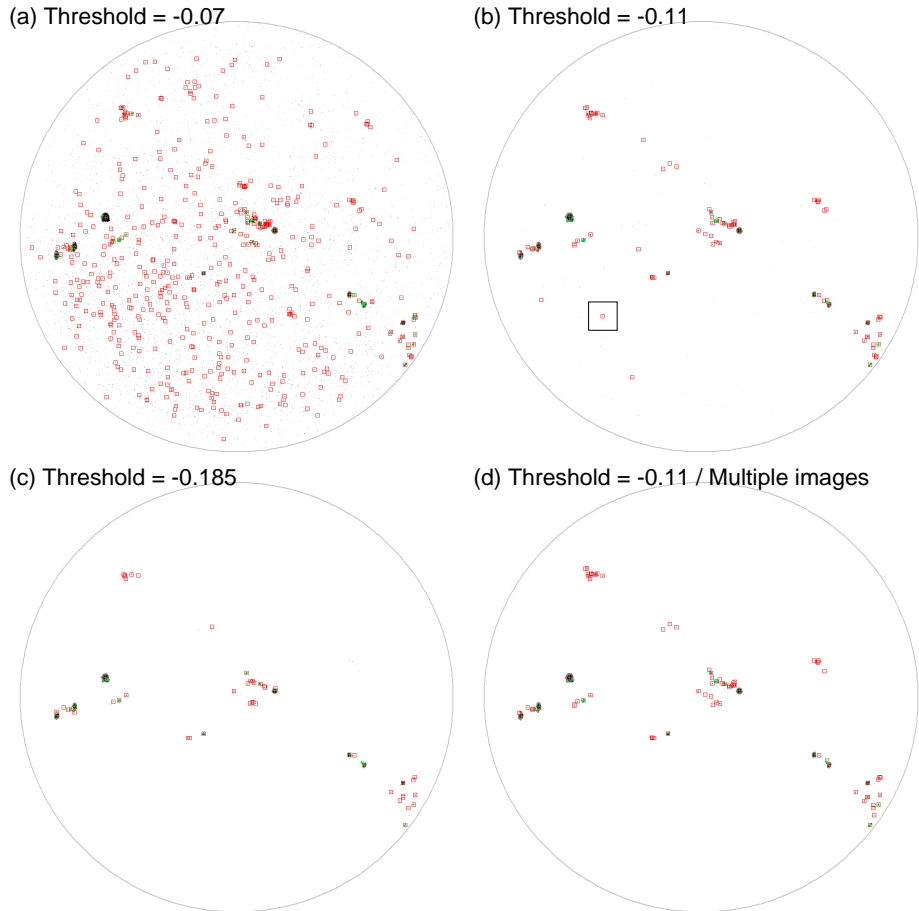}}
\caption{(a)--(c) Sunspots identified in the contrast image shown in Figure~\ref{fig:1}(d) using three different thresholds. (d) Sunspots identified using multiple images. Penumbrae and umbrae are represented with gray and black patches, whereas their center positions are indicated by red squares and green plus signs, respectively. The box with a black frame in panel (b) indicates one of the false sunspots that disappeared in multiple-image analysis; see Figure~\ref{fig:5}.}\label{fig:3}
\end{figure}

Figure~\ref{fig:3} depicts the spots identified with three different thresholds on the sample image shown in Figure~\ref{fig:1}(d). Positions of identified penumbrae and umbrae are indicated by red squares and green plus signs, respectively.
In the detection of sunspots shown in Figure~\ref{fig:3}, additional conditions were added to prevent false detections.
\begin{itemize}
\item A dark patch needs to cover a minimum number of pixels to be identified as a sunspot. Patches encompassing few pixels are possibly due to defects of detectors, and even if they are true sunspots, such small spots are considered difficult to be identified in visual observations.
\item Patches should constitute a real depression in the original image to be identified as a sunspot. A dent of the brightness near the limb, which is not a real depression and is mostly caused by the seeing effect, sometimes becomes a depression in the contrast image. Such patches are false sunspots and are excluded by referring to the original image.
\end{itemize}

Figure~\ref{fig:3}(b) shows the sunspots identified using the above rules with a threshold of $-0.11$ for penumbrae (pixels darker than neighboring photosphere by at least 11 \% are identified as penumbrae) and $-0.38$ for umbrae. They are mostly consistent with the spots visually identified in the image shown in Figures~\ref{fig:1}(a) and 1(d). By contrast, Figure~\ref{fig:3}(a) presents results with a higher threshold ($-0.07$ for penumbrae and $-0.26$ for umbrae); many false spots that correspond to intergranular dark lanes can be observed. Figure~\ref{fig:3}(c) shows results with a lower threshold ($-0.185$ for penumbrae and $-0.41$ for umbrae); in this case, some small spots are dropped, while suspicious detections disappear. 

% Figure 4
\begin{figure} 
\centerline{\includegraphics[width=0.9\textwidth]{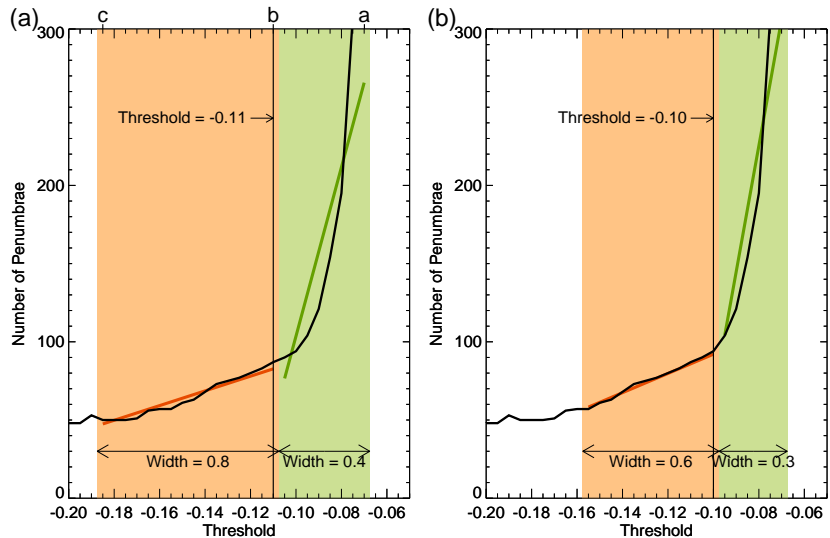}}
\caption{Relation between the threshold for penumbra and number of identified penumbrae along with the corresponding explanation concerning the determination of appropriate thresholds. The same curve showing this relation is plotted in panels (a) and (b); these panels depict two different examples of line fitting. The green line and green area indicate the fitting result and range of the data used for fitting in the ``high-threshold regime'', respectively. The orange line and orange area represent the fitting result and range of data used for fitting in the ``low-threshold regime''. The width of the range of data used for fitting in the ``low-threshold regime'' is 0.8 in panel (a) and 0.6 in panel (b). Labels a, b, and c at the top of panel (a) indicate the thresholds used to obtain the sunspot detection results depicted in Figures 3(a)--(c).}\label{fig:4}
\end{figure}

The results shown in Figure~\ref{fig:3} suggest that the best threshold is approximately $-0.11$. To determine the proper threshold for penumbrae, we derived the relation between the threshold value and number of detected penumbrae in the sample image. This is illustrated in Figure~\ref{fig:4}.

Figure~\ref{fig:4} indicates that there are two regimes across the turnoff point of the relation curve between the threshold and number of penumbrae.
For higher thresholds, the number of identified spots rapidly decreases with a decrease in the threshold. In this regime, many false sunspots are included in the identified spots, as depicted in Figure~\ref{fig:3}(a). For lower thresholds, the number of spots slowly decreases with a decrease in the threshold. In this regime, false spots are mostly excluded, but small true sunspots are gradually missed, as shown in Figure~\ref{fig:3}(c).
Therefore, a threshold around the turnoff point is expected to properly discriminate between false and true sunspots.

However, the brightness of false spots and that of true spots overlaps. The seeing effect sometimes produces small patches darker than some true spots (see an example in Section~\ref{s:2-4}). Therefore, it is impossible to divide them completely by using a threshold. The decision on how often the detection of false spots is allowed in automated detection depends on the strategy of the observers. Figure~\ref{fig:4}(a) shows that the threshold $-0.11$ is located somewhat lower than the turnoff point. Figure ~\ref{fig:4}(b) shows the threshold $-0.10$, which allows more frequent detection of false spots, located closer to the turnoff point. Such thresholds can be easily determined by visually inspecting the relation between the threshold and number of spots, but they should be determined automatically.
The procedure to determine the threshold, which can reflect the observers' preference, is carried out as explained next.

If we set windows in low- and high-threshold regimes (areas depicted in orange and green in Figure~\ref{fig:4}), we can fit the curve representing the relation between the threshold value and number of identified penumbrae within the windows with two lines (green and orange lines in Figure~\ref{fig:4}). In Figure~\ref{fig:4} (a), the
width of the low-threshold regime (orange) is 0.8 and that of the high-threshold
regime (green) is 0.4. In this case, the slope of the fitted line in the high-threshold regime is ten times steeper than that of the low-threshold regime. Figure~\ref{fig:4} (b), the fitted lines in the windows with the widths of 0.6 and 0.3, which are placed across the threshold line of $-0.10$, gives 1:10 slopes.
This fact means that a smaller window width gives a higher threshold, if the ratio between the slopes is fixed. Therefore, the threshold can be controlled by selecting width of the fitting window. If an observer requires a stricter threshold to reduce false spot detections, the width should be increased. 
For the data on a different day, if we set the positions of the windows with the same widths (such as 0.8 and 0.4) such that the slope of the fitted line in the high-threshold regime is a certain times (such as ten times) that in the low-threshold regime, we can determine the appropriate threshold with the similar strictness for that day.
The upper end of the fitting range in the low-threshold window was then adopted as the optimal threshold.

In Figure 4, the slope in the low-threshold regime has a finite value. However, it becomes close to zero in case of no spots or very small number of them. In such cases, it is impossible to specify the windows that gives an appropriate threshold. Therefore, we set the lower limit of the slope in the low-threshold regime to 120 (in the case of Figure~\ref{fig:4}(a), it is 470) on the basis of the analysis of the relation between the threshold and the detected dark features for no-spot days.

As stated above, the allowable frequency of false-spot detection depends on the strategy of the observers. Our standard width setting for multiple-image analysis is 0.8, as in Figure~\ref{fig:4}(a). This setting provides a rather strict threshold, with which we can expect a low frequency of false detection, allowing misdetection of true spots to some extent. In drawing observations, poorly-skilled observers or poor performance of instruments provide fewer sunspots; they are compensated for with the $k$-coefficient or personal coefficient of the observer. The reduction in the number of sunspots with strict thresholds in automated detection can be considered similar to individual differences in drawing observations. However, the detection of false spots is a serious problem because a single small isolated false spot incorrectly increases the relative sunspot number by 11. The detection of false spots is rare in drawing observations; therefore, it should also be infrequent in automated detection.

\subsection{Determination of True Spots Based on Multiple Images}\label{s:2-4} 

Given that small sunspots and false dark spots caused by the seeing effect in a single image cannot be distinguished completely, as described above, we included a spot identification process using multiple images in the standard procedures for spot detection.

The process is as follows. First, some of the images (for instance, the five best ones) are chosen from a series of images. Next, sunspot detection on the image exhibiting the best seeing condition among the selected images is conducted. The result may include false spots. Sunspot detection is then performed for the rest of the images. If a dark feature detected in the first image appears only in a small number of images (for instance, two images out of five or fewer), it is judged to be a false spot. Small true spots are also affected by the seeing effect, and they might not appear in the first image while appearing in other images. Such spots are considered difficult to be visually identified. Therefore, we allow their detection to be missed.

% Figure 5
\begin{figure} 
\centerline{\includegraphics[width=0.7\textwidth]{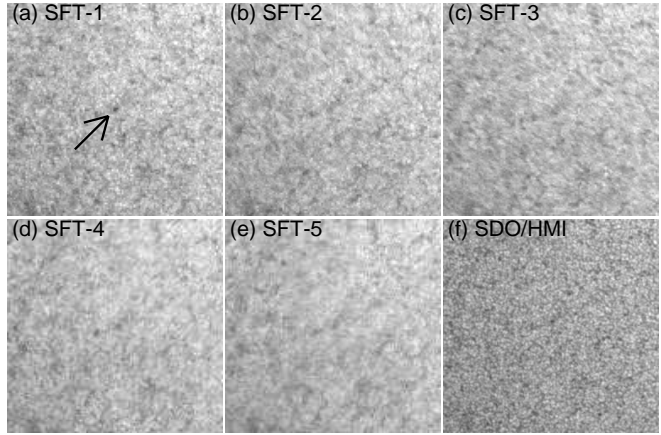}}
\caption{One of the false detected spots marked with a box in Figure~\ref{fig:3}(b). The field of view covers $2'.1 \times 2'.1$. Panels (a)--(e) show the corresponding portion in the five images used for multiple-image analysis. An arrow in panel (a) indicates a dark feature identified as a sunspot in Figure~\ref{fig:3}(b). Panel (f) depicts the same area in the white-light image taken by the SDO/HMI.}\label{fig:5}
\end{figure}

% Figure 6
\begin{figure} 
\centerline{\includegraphics[width=1.\textwidth]{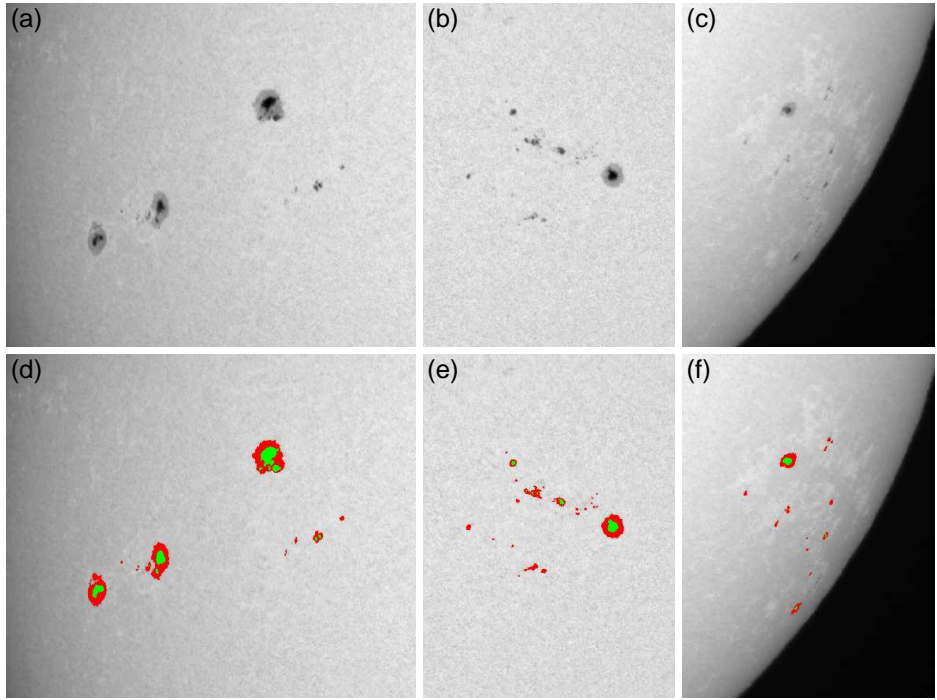}}
\caption{Sunspot detection results for some portions in the white-light image shown in Figure~\ref{fig:1}(a). Panels (a)--(c) depict enlarged views of the gray boxes in Figure~\ref{fig:1}(a), and in panels (d)--(f), penumbrae and umbrae, which are detected using multiple images, are indicated with red and green patches, respectively.}\label{fig:6}
\end{figure}

The original image shown in Figure~\ref{fig:1}(a) is the best image among 30 consecutive images; Figure~\ref{fig:3}(b) depicts the sunspot detection results with a properly chosen threshold. We processed four additional images selected from the aforementioned 30 images. Finally, spots judged to be true using the five images are shown in Figure~\ref{fig:3}(d). Some small spots in Figure~\ref{fig:3}(b) disappear in Figure~\ref{fig:3}(d). Although the spots in Figure~\ref{fig:3}(b) were identified using a rather strict threshold to reduce false spots, those in Figure~\ref{fig:3}(b) still include some false spots.
Figure~\ref{fig:5} shows enlargements of the portion including one of the false spots (indicated in Figure~\ref{fig:3}(b) with a box) for the five images, along with an image of the same portion taken with the SDO/HMI approximately at the same time. Note a dark spot (indicated with an arrow) in Figure~\ref{fig:5}(a). This was identified as a spot in Figure~\ref{fig:3}(b), but none of the remaining images show this spot clearly. This is a dark portion in the granulation, and instantaneously resembles a sunspot owing to the seeing effect.

A comparison of Figures~\ref{fig:3}(b) and \ref{fig:3}(d) indicates that when processing a single image alone, a lower threshold should be applied to reduce the number of false spots and minimize their effect on the relative sunspot number. This causes a reduction in the detection of small, true spots. By contrast, for multiple-image analysis, the detection of false spots in a single image is allowable to some extent because they can be excluded from the final result. Sunspot detection using multiple images is effective in detecting small spots with high reliability.

Figure~\ref{fig:6} shows some portions of the original image including major sunspots (panels (a)--(c), whose positions are indicated in Figure~\ref{fig:1}(a)) together with the results of automated detection using multiple images (panels (d)--(f), in which identified penumbrae and umbrae are represented with red and green patches).

\section{Results of Automated Detection }\label{s:3} 

\subsection{Data Used for Automated Detection and Performance Verification}\label{s:3-1} 

We tested the automated detection method described above by applying it to digital white-light observations using multiple images in all cases. The specifications of these instruments are summarized in Table~\ref{tbl:1}.

% Table 1
\begin{table}
\caption{Specifications of the imaging observations used for sunspot detection, visual drawing observations, and SDO/HMI}\label{tbl:1}
\begin{tabular}{lllrll}     
\hline
Observer, instrument, & Telescope & No. of pixels & Pixel & Bit & Selected/ total \\
\quad location & aperture & & scale & depth & no. of images \\
\hline
Solar Flare Telescope (SFT) & 12.5 cm & 2080$\times$2080 & $1''.00$ & 12 & 5/30 \\
\quad Mitaka, Tokyo, Japan \\
Kawaguchi Science Museum (KSM) & 10 cm & 2048$\times$2048 & $1''.18$ & 12 & 5/10 \\
\quad Kawaguchi, Saitama, Japan \\
S. Morita (SM) & 10.2 cm & 2208$\times$2208 & $0''.96$ & 8(jpeg) & 5/400 \\
\quad Moriyama, Shiga, Japan \\
\hline
Specola Solare Ticinese (SST) & 8 cm \\
\quad Locarno, Switzerland \\
Kwasan Observatory (KO) & 11.5 cm \\
\quad Kyoto, Japan \\
\hline
SDO/HMI & 14 cm & 4096$\times$4096 & $0''.5$ \\
\hline
\end{tabular}
\end{table}

The Solar Flare Telescope (SFT) is a set of synoptic observing instrument that includes imagers for various wavelengths and a spectropolarimeter for infrared wavelengths \citep{1995PASJ...47...81S, 2018PASJ...70...58S, 2020JSWSC..10...41H}. Broadband continuum images with the wavelength centered at 530 nm and a width of 50 nm are taken regularly with a 15-cm refractor (diaphragmed to 12.5 cm). Single-image acquisitions are performed every 5 min, and a series of 30 images is taken a few times per day. A set of 30 images is taken within 3 s. One of these datasets acquired per day is used for automated detection of sunspots. The acquisition of continuum images started in 2012. The five images featuring the highest quality were extracted from one of the sets of 30 images and used for detection of sunspots using multiple-image analysis. Concerning the standard parameters, we adopted 3 pixels for the minimum area of sunspots, 0.8 for the width of the fitting window, and 3 images out of 5 for the minimum number of images to identify sunspots. The results of sunspot detection presented in Section~\ref{s:2} were obtained using these parameters.

The Kawaguchi Science Museum (hereafter referred as KSM) is carrying out advanced solar observations; current white-light imaging observations began in 2011. They take a series of 10 images several times per day. For sunspot detection, five high-quality images are selected and processed with approximately the same parameters as those of the SFT. The images are contaminated with smears by the CCD detector; therefore, preprocessing is needed for removing these smears before sunspot detection.

Furthermore, we processed images taken by an amateur observer, Mr. S. Morita (hereafter referred as SM). He takes white-light images of the Sun and has been providing us with data since 2021. For sunspot detection, five high-quality images are selected and processed with the same parameters as those of the SFT except for the minimum area of sunspots, which is increased to 4 pixels because of fine sampling.

The results of automated detection using these data should be compared with other reliable data to verify the performance of the assessed method.
Generally, such results are compared with ground-truth data (correct results of sunspot detection). If visual observations are simultaneously carried out with digital image acquisitions under the same seeing conditions, they can be considered correct results. However, no such observations were made. Sunspots identified on digital images by visual inspection can be compared with the results of automated detection, but dark spots caused by the seeing effect on digital images will be misidentified as true spots, both by visual inspection and automated detection. Consequently, there is no perfectly correct results for sunspot detection.

Therefore, we used the results of drawing observations and white-light images of the SDO/HMI as reference data for comparison; they are also listed in Table~\ref{tbl:1}.

The drawing observations were made by the Specola Solare Ticinese and Kwasan Observatory of Kyoto University.
The Specola Solare Ticinese (hereafter referred as SST) has been conducting hand-drawn sketch observations of the Sun since 1957, and these observations constitute the reference data of the international sunspot number derived by the SILSO. Note that there is an eight-hour time difference between SST and imaging observations.
The Kwasan Observatory (hereafter referred as KO) began conducting drawing observations in 2004. Although the imaging observation sites and KO are located in Japan, their observation times are not the same.

Sunspots detected by the drawing observations were divided into groups to derive relative sunspot numbers. To compare the sunspot groups captured by imaging observations with those by drawing observations, we carried out grouping of sunspots for SFT and SM data, which we use for detailed comparison. Generally the method of sunspot grouping depends on the observers. Therefore, we carried out the sunspot grouping for SFT and SM data in two ways, one following SST and another following KO for respective comparisons.

The SDO/HMI takes continuum images with diffraction limit resolution every 45 s in space. A HMI image taken approximately at the same time as each of the images used for automated detection can be found. Therefore, they can be used as high-resolution reference images.

Using these reference data, we verified the results of automated spot detection by checking the following points, according to the criteria explained in Section~\ref{s:2}.
\begin{itemize}
\item Whether the number of sunspots detected with the automated method is comparable to that of visual observations.
\item Whether the frequency of false detection of sunspots is as low as in visual observations.
\item Whether no bias exists in missed detections of sunspots, as in visual observations.
\end{itemize}

\subsection{Comparison of the Number of Detected Sunspots}\label{s:3-2} 

We compared the number of sunspots derived from automated detection and drawing observations for 2014 and 2021, which are close to the solar maximum and minimum, respectively, from some points of view.

The first one is total number of sunspots. It is often represented by the number of umbrae in the drawing observations, and penumbrae without umbrae are not counted. However, as presented in Figure~\ref{fig:6}, small spots, which will be classified as umbrae without penumbrae in visual observations, are identified as penumbrae without umbrae by automated detection. This is an effect of the scattered light. Therefore, we calculate the total number of sunspots as the sum of the number of penumbrae without umbrae and number of umbrae from the results of automated detection.

The SST basically uses the weighted sunspot number, taking the size of the spot and concomitance of the penumbra into account. However, they provide the number of individual umbrae as well; we used this number for comparison.
% Figure 7
\begin{figure} 
\centerline{\includegraphics[width=0.8\textwidth]{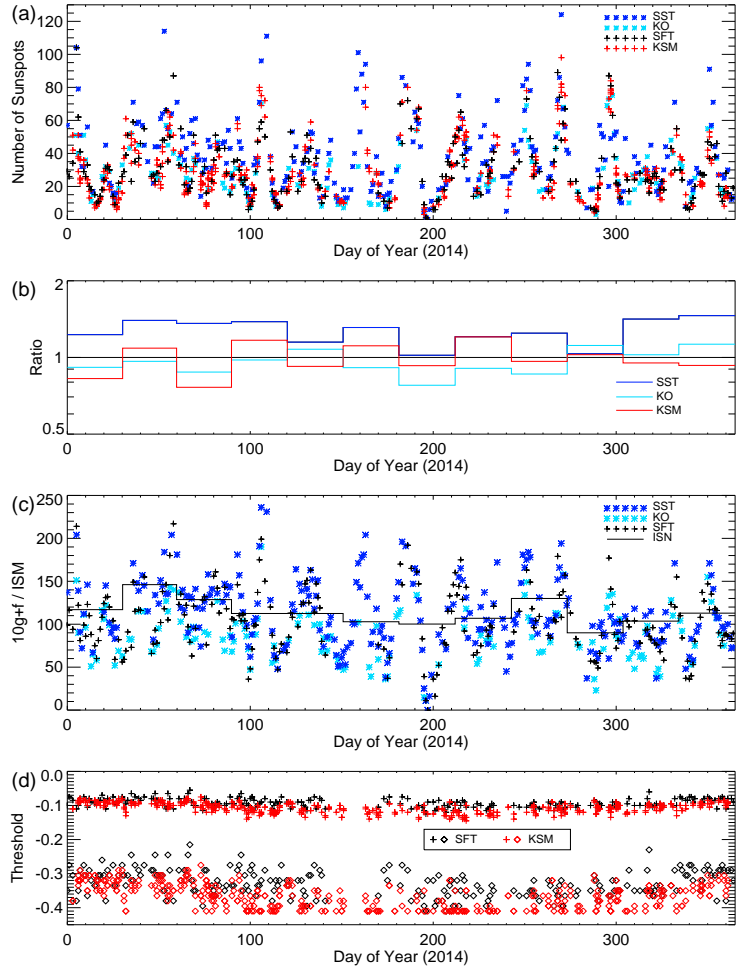}}
\caption{Comparison of detection results based on digital white-light images and visual observations for 2014. (a) Daily number of sunspots obtained by automated detection for the SFT and KSM data and those by SST and KO visual observations. (b) Ratio of the number of sunspots detected with KSM, SST, and KO observations with respect to those by the SFT each month. (c) Daily $10g+f$ values observed by SST, KO, and SFT. The monthly-mean international sunspot numbers are also presented. (d) Thresholds to detect penumbrae (plus signs) and umbrae (diamonds) used in automated detection for SFT and KSM data.}\label{fig:7}
\end{figure}

% Figure 8
\begin{figure} 
\centerline{\includegraphics[width=0.8\textwidth]{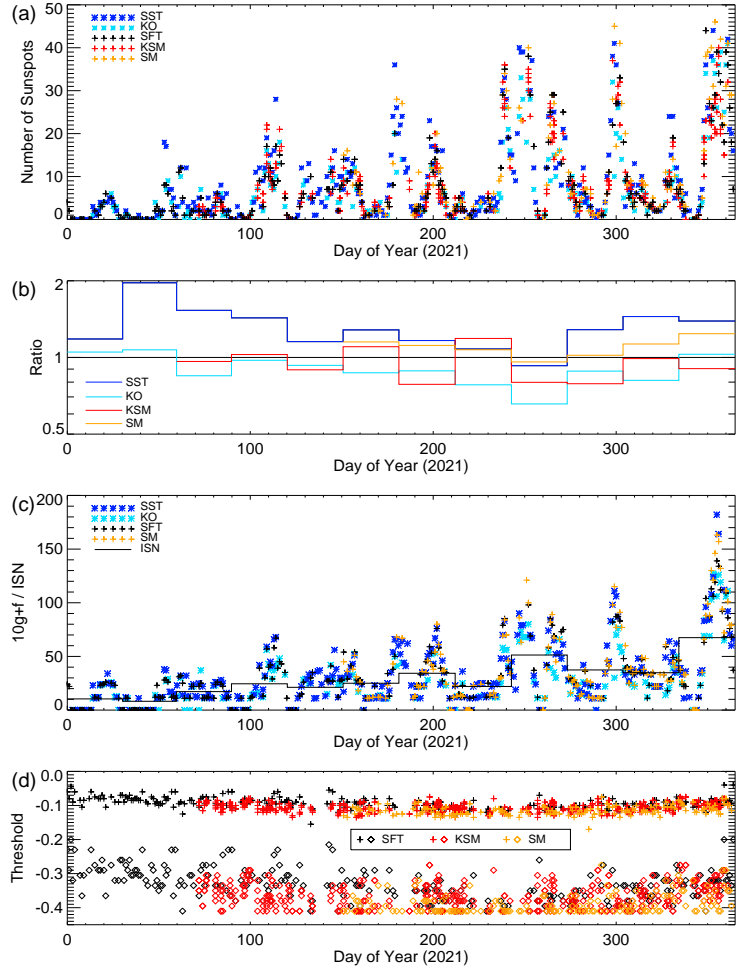}}
\caption{Comparison of the detection results based on digital white-light images and results of visual observations for 2021. The panels were plotted in the same way as in Figure~\ref{fig:7} except for the fact that automated detection results by SM were added.}\label{fig:8}
\end{figure}

Figures~\ref{fig:7}(a) and \ref{fig:8}(a) show the total number of sunspots derived from automated detections and drawing observations for 2014 and 2021, respectively. The variations in the number of sunspots obtained with various observations were similar, but there was a significant scatter. Therefore, we present the ratios of total sunspot numbers with respect to the SFT in Figures~\ref{fig:7}(b) and \ref{fig:8}(b). Each ratio is calculated from the total number of sunspots on the common observation days in a month for the SFT and another instrument. We found a tendency for the SST to capture more sunspots than others, and that the KO and KSM captured slightly fewer spots than the SFT.

The second one is the $10g+f$ values, because the number of sunspots ($f$) are used with the number of groups ($g$) to represent the activity of the Sun with the form of $10g+f$. In Figures~\ref{fig:7}(c) and \ref{fig:8}(c), we present the $10g+f$ values for SST, KO, SFT, and SM observations. The monthly mean international sunspot number is also presented in Figures~\ref{fig:7}(c) and \ref{fig:8}(c). It is confirmed that the $10g+f$ values follow the international sunspot number well.

Table~\ref{tbl:2} presents quantitative comparison of the number of sunspots detected by various observations. In addition to the ratio of the annual total number of sunspots and the ratio of annual $10g+f$ values, the ratio of the groupwise number of sunspots is presented.

For the ratio of the annual total number of sunspots of each instrument to that of the SFT for the common observation days of both instruments, the SST ratios were the highest. This indicates the advantage of drawing observations by skilled observers. The ratios of the total number of sunspots show that all SFT, KSM, and SM observations captured more sunspots than those of KO. Automated detection uses images with a scale of approximately 1$'' {\rm pixel}^{-1}$, and the resolution in such images are not diffraction limited. The order of the ratio of the total number of sunspots of the SM, SFT, and KSM observations corresponds to that of the pixel scale (see Table~\ref{tbl:1}). This result suggests that automated detection using images taken with 10-cm class telescopes with sufficient spatial sampling provides comparable results to those of the SST.

The ratio of annual $10g+f$ values for the common observation days of two instruments shows a similar tendency to that of the ratio of the total number of sunspots, but it is affected by the capacity of capturing sunspot groups. In the case of SST and SM, the ratios of $10g+f$ are closer to unity than the ratios of the total sunspot number. This means that SFT, SST, and SM have a similar capacity to capture groups, while there are systematic differences in the capacity to detect sunspots. In contrast, the of $10g+f$ ratios of KO is further away from unity than the ratios of the total sunspot number. This is because KO often missed to capture sunspot groups.  The capacity of capturing sunspot groups is discussed in Section \ref{s:3-3}.

The ratio of the groupwise number of sunspots shows the mean ratio of the number of sunspots in individual groups commonly observed by SFT and other instruments. To check the capacity of detecting sunspots in individual groups, one-to-one comparison between the sunspots detected in each group by different instruments is the best way. However, because of the time difference between the observations, such a comparison is difficult. Therefore, we compared the number of sunspots in individual groups detected by different observations statistically. The grouping of SFT and SM sunspots was done according to that of SST for the comparison with the SST data and according to that of KO for the comparison with the KO data.

The ratios of the groupwise number of sunspots presented in Table~\ref{tbl:2} are similar to those of annual number of sunspots, and it is confirmed that the observed number of sunspots in groups is basically consistent with the total number of sunspots. However, it is noteworthy that the ratios of the groupwise number of sunspots of KO is higher than those of total number of sunspots. This again indicates that KO often missed to capture sunspot groups, while the KO's capacity to detect sunspots in individual groups is comparable to that of SFT.

% Table 2
\begin{table}
\caption{Ratio of the number of detected sunspots with respect to SFT observations}\label{tbl:2}
\begin{tabular}{llrrrrr}     
\hline
 & & SFT &KSM\tabnote{KSM obtains data a few times in a day; 357 (2014) and 394 (2021) observations are used to derive averages.} &SM &KO &SST \\
2014 & No. of days\tabnote{Number of days in which observations were made with both instruments.} & (219)\tabnote{Total number of SFT observation days} &153 & -- & 104 & 144 \\
 & Ratio (Annual number of sunspots) & (1) & 0.98 & -- & 0.96 & 1.26 \\
& Ratio (Annual 10$g+f$) & (1) & -- & -- & 0.87 & 1.09 \\
& Ratio (Groupwise number of sunspots) & (1) &-- & -- & 1.02 & 1.22 \\
2021 & No. of days & (238) & 131 & 102 & 170 & 177 \\
 & Ratio (Annual number of sunspots) & (1) & 0.93 & 1.13 & 0.90 & 1.28 \\
& Ratio (Annual 10$g+f$) & (1) & -- & 1.04 & 0.87 & 1.07 \\
& Ratio (Groupwise number of sunspots) & (1) & -- & 1.10 & 0.99 & 1.22 \\
\hline
\end{tabular}
\end{table} 

As explained in Section~\ref{s:2-3}, determining the brightness threshold is important for sunspot detection. Figures~\ref{fig:7}(d) and \ref{fig:8}(d) show the derived thresholds for the penumbra and umbra of the SFT, KSM, and SM observations; note that they exhibit some scatter. 
To check the reliability of the threshold determination, we carried out automated sunspot detection using HMI white-light images taken almost simultaneously with SFT observations in 2014. The standard deviation of the threshold for the penumbra of the HMI data taken without seeing effect was 0.005 (the average is $-0.19$), while it was 0.013 for the SFT data and 0.014 for the KSM data. The fact that the threshold for the HMI data exhibits only slight scatter confirms the stability of threshold determination in automated detection. The scatter observed in Figures~\ref{fig:7}(d) and \ref{fig:8}(d) is supposed to be due to the seeing effect. Note the slight decrease of the thresholds around 200--250 days of year or summer season in Figures~\ref{fig:7}(d) and \ref{fig:8}(d). This is consistent with the general trend of the seeing conditions at the observation sites of the SFT, KSM, and SM. This also confirms that the thresholds were determined appropriately. 

\subsection{False Spot Detection}\label{s:3-3} 

Next, we checked whether false spots were included in the features detected by the automated method. We compared the observational results of the SFT and SM, which are good at detecting small spots, with continuum images obtained with the HMI visually. 
It is difficult to compare individual spots in groups containing many spots; however, groups consisting of one or a few small sunspots can be easily checked. If we cannot find a group on a HMI image corresponding to that detected by the automated method, it is probably false. Drawing data from the SST and KO obtained on the same days as SFT or SM observations were also checked. 

% Figure 9
\begin{figure} 
\centerline{\includegraphics[width=0.6\textwidth]{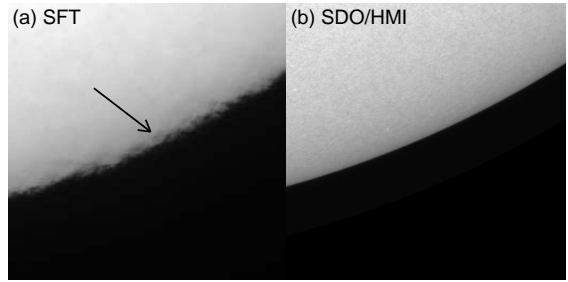}}
\caption{False detections of dark features in an image taken with the SFT on 2021 January 19. A dent pointed by an arrow was misunderstood as a sunspot in the automated detection even using multiple images.}\label{fig:9}
\end{figure}

SM data for 143 days in 2021 showed no false sunspots.
However, we found an isolated false spot within 457-day observations in 2014 and 2021 for the SFT data on 2021 January 19, as shown in Figure~\ref{fig:9}(a). Figure~\ref{fig:9}(b) presents an image nearly simultaneously taken with SDO/HMI that shows no sunspots. It appeared at the limb on the image taken under poor seeing conditions. Blurring caused by the seeing occasionally produces dents near the limb; if a dent appears at a similar position repeatedly, it is misunderstood as a sunspot. Such a false spot at the limb can be easily excluded by visual inspection. Observations under unusually poor seeing conditions, such as those shown in Figure~\ref{fig:9}, are not necessarily adequate for inclusion in the statistical analysis of the sunspot number.

In automated detection, dust on images may be identified as sunspots. Therefore, it is desirable to conduct a visual inspection of the detection results in any case.

Drawing observations are not free from suspicious sunspots. 
We found a group in SST results that cannot be recognized in HMI images after inspection of 322-day observations in 2014 and 2021, when SFT or SM observations were also performed and the sunspot number for at least one of the observations was not 0. Checking the KO results of the 277-day observations, we found a group that cannot be recognized in HMI images. It is difficult to consider that these are true spots, and therefore, the frequency of false group detection using the automated method is not higher than that of drawing observations. 

The threshold of sunspot detection discussed above is controlled by the window width of the ``lower-threshold regime'' shown in Figure~\ref{fig:4}. The aforementioned results for sunspot detection were obtained with a width of 0.8. For a smaller window width, the threshold increases, as shown in Figure~\ref{fig:4}(b), and fainter sunspots are  detected. For a width of 0.6, the number of spots increased by approximately 5 \% in the SFT data. The results with a width of 0.6 for 457 days show five additional false spot groups, which cannot be identified as sunspots in the HMI data. The SM data for 143 days processed with a width of 0.6 reveal three false spot groups. These false-spot groups are non-spot dark structures on the solar disk (e.g., intergranular lanes).

Some observers may think that the increase in the number of false spot groups is not significant compared to the 5 \% increase in the number of detected spots. We adopted a safe-side threshold at which the false-detection frequency is as low as that of the drawing observations. However, the decision regarding the strictness of the threshold is decided by the observers. Our method is notably flexible in this regard.

\subsection{Comparison of Misdetections between the Automated Method and Drawing Observations}\label{s:3-4} 

As mentioned above, appropriate thresholds mostly prevent false detection of sunspots. However, small true spots may be missed with such thresholds. Misdetections themselves are not a serious problem as long as they remain small in number. However, there should be no bias in the distribution of missed detections of spots in the automated method to achieve a detection performance comparable to that of visual observations. 

% Table 3
\begin{table}
\caption{Comparison of number of sunspot groups captured by automated detection and visual observations}\label{tbl:3}
\begin{tabular}{lrrrr}     
\hline
Automated detection & \multicolumn{2}{c}{SFT} & \multicolumn{2}{c}{SM} \\
Visual observation & KO & SST & KO & SST \\
\hline
Identified by at least one of the observations & 1090 & 1587 & 283 & 314 \\
Identified only by automated detection & 159 & 114 & 57 & 33 \\
Identified only by visual observation & 21 & 135 & 3 & 28 \\
\hline
\end{tabular}
\end{table}

To verify this no-bias condition, we compared the sunspot groups detected by the automated method (SFT and SM) and those detected by visual observations (SST and KO) in 2014 and 2021.
The results are summarized in Table~\ref{tbl:3}. While most of the groups were captured by both automated detection and visual observation, a certain number of groups were captured by only one of the two observation modalities. The comparison between SFT/SM and KO indicates that SFT/SM observations missed only a small number of groups. This is consistent with the fact that SFT/SM captured more sunspots than KO, as shown in Table~\ref{tbl:2}.

% Figure 10
\begin{figure} 
\centerline{\includegraphics[width=0.7\textwidth]{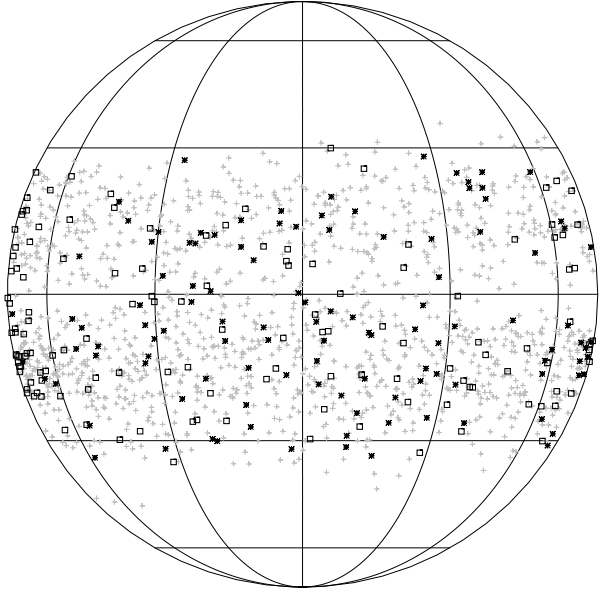}}
\caption{Sunspot groups captured by SFT or SST observations in 2014 and 2021. Gray plus signs show the position of the groups captured by both SFT and SST observations. Stars and squares indicate the position of the groups captured only by the SFT and those captured only by the SST. The longitude and latitude lines, drawn every 30 degrees, are depicted for reference; they do not represent the heliographic coordinates of the groups.}\label{fig:10}
\end{figure}

By contrast, the comparison between SFT/SM and SST indicates that more groups were missed by both automated detection and visual observation because SST observation was carried out approximately eight hours later than the corresponding SFT and SM observations owing to the time difference between the observation sites. Many groups are supposed to appear and disappear during a time interval, and they are identified as only one of the observations, as well as missed detections of spot groups. 
Given that the spot groups detected by the SFT or SST include larger samples, we show the distribution of the spot groups detected by them in Figure~\ref{fig:10}. The groups detected by both SFT and SST observations are represented with gray plus symbols, and those detected only by SFT and only by SST are represented with black stars and square symbols, respectively. Because of the rotation of the Sun during the time difference, there is a concentration of groups detected only by the SST near the east limb and another detected only by the SFT near the west limb. Except for them, the groups detected either from SFT or SST observations distribute with no significant bias. This means that the center-to-limb variation of the detectability in automated detection is similar to that of SST drawing observations.

In summary, the total number of spots detected by the automated method is not smaller than that detected by visual observation, the frequency of false detection in the automated method is not high, and there is no bias in the missed detections of spots in the automated method. Therefore, we can conclude that the automated detection method achieves a level comparable to that of drawing observations.

\section{Concluding Remarks}\label{s:4} 

We developed a new automated sunspot detection method using digital white-light images. Its performance is similar to that of visual observation and can be used as an alternative. The key requirement for such a method is that the total number of detected sunspots, false detection of spots, and missed detections of true spots should be comparable to those of visual drawing observations. 

To meet this requirement, we focused on the derivation of the quiet-disk component of the Sun, which is the reference for deriving the brightness depression of sunspots. We reproduced the disk component by deriving limb darkening and adjusting the shape of the disk distorted by the seeing effect. To correctly identify sunspots under various conditions, we determined the appropriate threshold using an adaptive method that inspects the relationship between the threshold and the number of detected sunspots. 
In addition, to prevent false detection of dark features instantaneously appearing by the seeing effect as sunspots, we added a function to process multiple images taken within a short time interval.

We applied this method to detect sunspots for the digital images taken with the Solar Flare Telecope, the Kawaguchi Science Museum, and by Mr. S. Morita, and compared the results with visual observations performed at the Specola Solare Ticinese and the Kwasan Observatory, and data obtained with the SDO/HMI to evaluate the performance of the automated detection method. From the comparison, we conclude that the aforementioned requirement is fulfilled.

The total number of detected sunspots is greater than that of the observations at the Kwasan Observatory and not much smaller than that from the Specola Solare Ticinese. The imaging observations were performed with 10--12.5 cm telescopes, but the images were not diffraction-limited owing to limited pixel sampling. Automated sunspot detection, which is comparable to the observation by the Specola Solare Ticinese, is considered to be achieved with finer pixel sampling. 

The automated method is flexible enough to process a variety of digital data. Therefore, it can be used by various observers such as public observatories and amateurs. The Kawaguchi Science Museum, which formerly conducted drawing observations but is now carrying out only white-light image acquisitions, can currently provide sunspot counting results using the automated detection method. 
This application of the proposed detection method as an alternative to hand-drawn visual observations for sunspot counting meets the aim of this study. In the future, it is necessary to check long-term consistency between such observations and the international sunspot number, which are basically determined by the drawing observations, using the data over more than a solar cycle.

% Acknowledgements
\begin{acks}
We thank the Kawaguchi Science Museum and Mr. S. Morita, who provided digital white-light images of the sun. The Specola Solare Ticinese and Kwasan Observatory kindly permitted us to use their visual sunspot observation data for our analysis.
HMI data used in this study were courtesy of NASA/SDO and the HMI science team.
SDO is a mission for NASA's Living With a Star program. The source of the international sunspot number is WDC-SILSO, Royal Observatory of Belgium, Brussels.
\end{acks}

%% Available additional data environments:
%% required: authorcontribution, fundinginformation, dataavailability
%% optional: materialsavailability, codeavailability
% \begin{authorcontribution}
%
% \end{authorcontribution}
%
% \begin{fundinginformation}
%
% \end{fundinginformation}
%
% \begin{dataavailability}
%
% \end{dataavailability}
%
% \begin{ethics}
% \begin{conflict}
%
% \end{conflict}
% \end{ethics}

%%% %%%%%%%%%%%%%%%%%%%%%%%%%%%%%%%%%%%%%%%%%%%%%%%%%%%%%%%%%%%
%% Bibliography
%
% Using BibTeX
%
%\bibliographystyle{spr-mp-sola}
%\bibliography{hanaoka_R1}  
%
% Without BibTeX 
% \begin{thebibliography}{}
% \bibitem[\protect\citeauthoryear{Author}{Year}]{key}
%   <bibliographical entry>
%
% \bibitem[\protect\citeauthoryear{}{}]{}
%   
%  
% \end{thebibliography}

\end{article} 
\end{document}